\documentclass{entcs}
\usepackage{prentcsmacro}
\usepackage{graphicx}
\usepackage{color}
\def\lastname{Boutillier, Cristescu}
\begin{document}
\begin{frontmatter}
  \title{When rule-based models need to count}
  \author{Pierre Boutillier\thanksref{pierreemail}}
  \author{Ioana Cristescu\thanksref{ioanaemail}}
  \address{Fontana Lab\\Harvard Medical School\\
    Boston, USA} \thanks[pierreemail]{Email:
    \href{mailto:Pierre\_Boutillier@hms.harvard.edu} {\texttt{\normalshape
        Pierre\_Boutillier@hms.harvard.edu}}} \thanks[ioanaemail]{Email:
    \href{mailto:Ioana\_Cristescu@hms.harvard.edu} {\texttt{\normalshape
        Ioana\_Cristescu@hms.harvard.edu}}}
\begin{abstract}
Rule-based modelers dislike direct enumeration of cases when more efficient means of enumeration are available.
We present an extension of the Kappa language which
attaches to agents a notion of \emph{level}. We detail two encodings
that are more concise than the former practice.
\end{abstract}

\end{frontmatter}

Rule-based languages are a well-established framework for modeling protein-protein interactions.

Kappa~\cite{DanosFFHK08,kasim} is a rule-based language relying on
\emph{site-graphs}. The nodes of site-graphs are called
\emph{agents}. Agents interact by binding/unbinding through
\emph{sites}. Sites are binding resources, each site is part of at
most one edge.

A model in Kappa consists of a set of graph rewrite rules with rates.  A rule
describes a potential interaction given a context. Rates represent
probability to fire.

In a biological context, it is often the case that a notion of
internal state (such as \texttt{active}, \texttt{methilated}, \dots)
is required in order to describe possible interactions. In Kappa, sites
are equipped with internal states, facilitating the modeling efforts
of the user.  However, as shown in the following sections a more systematic encoding of internal states is possible.

Another common practice is to attach a \emph{level} to agents and make
an interaction sensitive on the level of its participating agents.

We propose here a language extension to store, test and change levels explicitly. Moreover, we present an encoding of levels
that induce a linear (in number of levels) blow up of the number of rules.  This is in contrast to previous encodings, which induce an exponential blow up in the number of rules.

\section{When enumeration is necessary}
The following motivating example~\cite{golden2007integrating} demonstrates a typical problem in which levels are necessary:
\begin{quotation}
KaiC proteins have 6 independent phosphorylation
sites. (De)phosphorylation of every site is independent. The more
sites are phosphorylated, the bigger the probability that KaiC binds
KaiA is.
\end{quotation}

A typical way to deal with this example consist in explicitly encoding rules for the internal states of the sites of interest. However, doing so induces
an exponential blow-up in the number of rules (in the number of levels).

The BNGL~\cite{bngl} language introduces a notion of
\emph{indistinguishable} sites. i.e. one can define an agent with $n$
sites that all have the same name. Consequently, a single rule to
specify that $k$ sites (out of $n$) are phosphorylated is
enough. Moreover, the number of species is also reduced. Still, there
are exponentially many ways to go from one level to another and
enumeration is necessary to faithfully respect the dynamics of the system.

We now show that adding a syntactic layer to Kappa that offers support for counting
will avoid the explosion in the number of rules.

\section{Encoding counters}

In Kappa, agents are typed and their \emph{signature} is given.  In
the extension we are presenting, counters are part of an agent's signature and
the upper bound of the counter has to be specified.

Extended rules can test counter values (``equals'' and ``bigger
than''). They can also modify their value (increase, decrease or
assign a new value). Lastly, levels can be used in the rates of the
rules.

We now present two possible ways of encoding counters.

\subsection{Unary numbers}

\begin{minipage}{.6\textwidth}
We define a new agent type \verb'succ' with 2 sites \verb'p' and
\verb'n'. A chains of $k+1$ \verb'succ' agents denotes a counter with
value $k$. Agents equipped with counters are encoded as agents with an
extra site for each counter, which is used to bind to a chain
\end{minipage}
\begin{minipage}{.39\textwidth}
\centering
\includegraphics[scale=0.5]{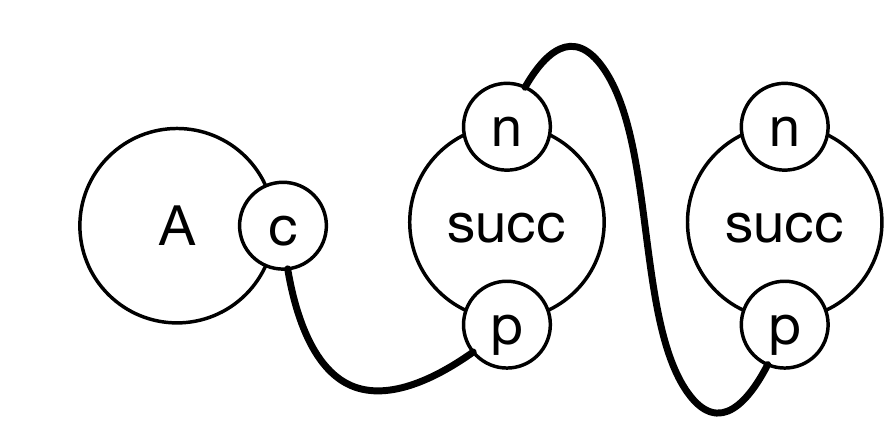}
\end{minipage}
of \verb'succ' agents. For example, in the figure on the right, an agent
\verb'A' has a counter \verb'c' set to $1$\footnote{Representing level
  $0$ by a chain of length $1$ means that we do not have to make a special case for
  it in the rules!}.

Testing whether a counter is equal to $k$ consists in checking whether
there is a chain of $k+1$ \verb'succ' agents. A greater than $k$
test checks whether there is a chain $k+1$ \verb'succ' agents, where
the site \verb'n' of the last \verb'succ' agent does not matter.

\hspace*{-\parindent}%
\begin{minipage}{.69\textwidth}
A counter is incremented by adding \verb'succ' agents between the
agent and its \verb'succ' chain. The rule depicted on the right
increments \verb'c' of \verb'A' by one.  Removing the beginning of the
\verb'succ' chain decreases the level.

It is important to stress that counter modifications are independent of its value as the
encoding only manipulates the beginning of a \verb'succ' chain.
\end{minipage}
\begin{minipage}{.30\textwidth}
\centering
\includegraphics[scale=0.5]{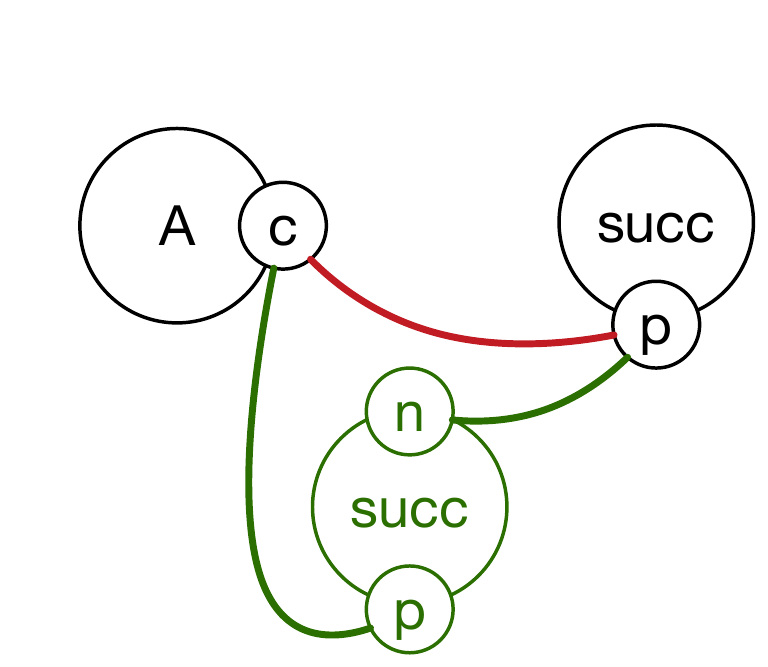}
\end{minipage}

In our encoding, a rule whose rate depends on levels is
expanded into several rules, one for each level. This operation
imposes the user-defined upper bound on the levels.

Creation of agents with levels also creates the necessary chain of
\verb'succ' to represent the levels.

Deletion is more problematic: the chain of \verb'succ' is disconnected
from the deleted agent, but not deleted. A possible solution is to collect free chains of \verb'succ' by using a rule that says
that \verb'succ' agents with their sites \verb'p' free are deleted ``at
infinite speed''.

\subsection{Ruler}
A second encoding allows tests ``smaller than'' in addition of
``equal'' and ``greater than'', but is also more verbose and increases
the size of states.

\hspace*{-\parindent}%
\begin{minipage}{.61\textwidth}
As before, counters are encoded as a chain of \verb'succ' agents. In
this encoding however, a \verb'succ' agent has 3 sites: \verb'p' and
\verb'n' to form chains but also \verb!a!, which is where the other
agents bind.
Every agent with a counter bounded by $n$ has attached a chain of
(always) $n$ \verb!succ!. The value of the counter is given by which
\verb!succ! agent it is bound to on site \verb'a'.  For example, in
the figure on the right, an agent \verb'A' has a counter \verb'c' set
to $1$.
\end{minipage}
\begin{minipage}{.38 \textwidth}
\includegraphics[scale=0.5]{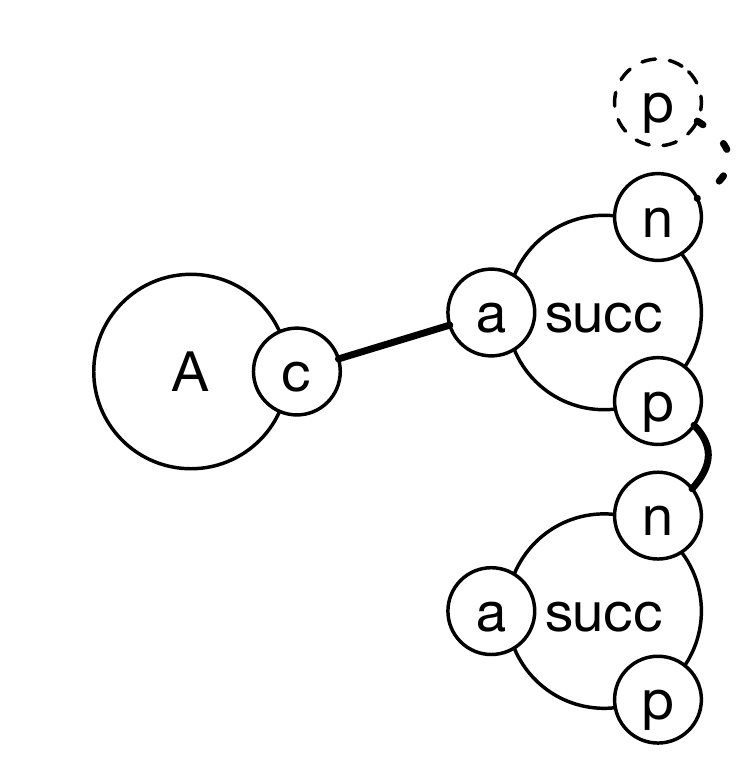}
\end{minipage}

Since site \verb'p' and \verb'n' are distinct, a direction can be
attached to the chain of \verb'succ'. The level is $0$ if the agent is
bound to the ``bottom'' of the chain, and it is $n$ if it is bound to the
``top''.

Incrementing/decrementing the levels are implemented by sliding on the
\verb'succ' chain.
Testing whether a level is equal or greater than $k$ consists of
inspecting the chain of \verb'succ' ``below'' the connection to the
agent. Whereas inspecting the length of the chain ``above'' the connection to
the agent enables to test if the level is smaller than a value $k$.

\section{Conclusions}

While fairly trivial, encoding counters in the Kappa simulator has
greatly simplified some of the models written by the Kappa team.
However, it comes with two drawbacks. First, there is a computational
cost (mainly in term of memory management) to synthetize and degrade
\verb'succ' agents. Secondly, the simulator works with a plain Kappa
model.  There is an implementation cost to go back from the plain Kappa
model to the user-written model, which is necessary when giving
feedback to the user (for example, dumping the current state of the
system).  As a result, we plan to implement a native treatment of the
counters in the Kappa simulator.

Another direction for future work is the static detection of overflow
for counters.  Currently, the maximal level of the counter has to be declared by
the user. Nevertheless, this means the rules have to be written such that they can never increase a level beyond the defined boundary. The level of the counter can be
checked 
by static analysis using
interval computation, but for the meantime we have simply added
watchdogs that dynamically raise the alarm if the site of a level becomes
free, or if a chain of \verb'succ' that is too long appears.



\bibliographystyle{plain}
\bibliography{cite.bib}

\end{document}